\documentclass[11pt]{JHEP3}
\usepackage{epsfig}

\def\spose#1{\hbox to 0pt{#1\hss}}
\def\ltapprox{\mathrel{\spose{\lower 3pt\hbox{$\mathchar"218$}}
 \raise 2.0pt\hbox{$\mathchar"13C$}}}
\def\gtapprox{\mathrel{\spose{\lower 3pt\hbox{$\mathchar"218$}}
 \raise 2.0pt\hbox{$\mathchar"13E$}}}

\def\sss{\scriptscriptstyle}

\preprint{SISSA-49/2003/EP}

\title{Neutralino relic density in supersymmetric GUTs\\ with no-scale boundary
conditions above the unification scale}
\author{Stefano Profumo \\
        Scuola Internazionale Superiore di Studi Avanzati \\
	Via Beirut 2-4, I-34014 Trieste, Italy \\
and Istituto Nazionale di Fisica Nucleare, Sezione di Trieste, 
I-34014 Trieste, Italy\\ 
	E-mail: \email{profumo@sissa.it} 
}

\date{\today}

\abstract{We investigate $SU(5)$ and $SO(10)$ GUTs with vanishing scalar
masses and trilinear scalar couplings at a scale higher than the unification scale.
The parameter space of the models, further constrained by $b$-$\tau$ Yukawa coupling unification, consists of a common gaugino mass and of $\tan\beta$. We analyze the low energy phenomenology, finding that $A$-pole annihilations of neutralinos and/or coannihilations with the lightest stau drive the relic density within the cosmologically preferred range in a significant region of the allowed parameter space. Implications for neutralino direct detection and for CERN LHC experiments are also discussed.}

\keywords{Supersymmetry Phenomenology, GUT, Dark Matter}

\begin{document}

\section{Introduction}

A critical issue concerning the supersymmetric extensions of the
Standard Model is associated with the mechanism of supersymmetry (SUSY)
breaking. The resulting pattern of soft SUSY breaking (SSB) terms, appearing
in the effective Lagrangian after integrating over the so-called
hidden sector, determines the low energy phenomenology. This
pattern is constrained both by theoretical (e.g. naturalness in
the Higgs sector) and phenomenological (e.g. flavor changing
neutral currents, FCNC) requirements. Nevertheless, the number
of {\em a priori} free high energy parameters in generic SUSY
breaking scenarios is uncomfortably very large.\\ There exist,
however, particular scenarios where one expects some of these
parameters to vanish. In the context of the so called {\em
no-scale models} \cite{noscale}, the scalar soft breaking masses vanish at some
high energy {\em boundary conditions} input scale $M_{\rm bc}$. Analogous boundary conditions arise in extra dimensional brane models with gauge mediated SUSY breaking \cite{gauginomediated}.
The low energy particle spectrum then depends, through renormalization group (RG) running, on the
non-vanishing input parameters of the theory, e.g. the gaugino
masses.\\
In ref.\cite{boundm0} it was
shown that requiring $m_0=0$ at $M_{\sss\rm GUT}\simeq 2\times 10^{16}\ {\rm GeV}$
is not compatible with low energy phenomenology. Nevertheless, if $M_{\sss\rm GUT}<M_{\rm bc}\lesssim M_{\rm
Pl}=2.4\times10^{18}\ {\rm GeV}$, GUT
interactions running may generate viable low energy spectra \cite{minimalgaugino}.\\
In this paper we investigate this possibility focusing on two simple GUTs, namely $SU(5)$ and general $SO(10)$. The requirement of Yukawa coupling unification at the GUT scale, together with all known phenomenological constraints, highly restricts the parameter space of the models under scrutiny. In the minimal setting to which we resort, once fixed the $M_{\rm bc}$ scale, only two parameters determine the particle spectrum, namely the common gaugino mass at $M_{\sss GUT}$ and the ratio of the two Higgs vevs, $\tan\beta$.
\\ 
We show in what follows that a
portion of the allowed parameter space is compatible with the
current data on the cold dark matter content of the Universe from
the WMAP survey \cite{wmap}. We also discuss the prospects for
neutralino direct detection \cite{directdetection}, and
argue that a large part of the cosmologically preferred region will be within 
reach of LHC \cite{reach}.

\section{Grand unified models with no-scale SSB scale higher than $M_{\sss\rm GUT}$}

No-scale boundary conditions naturally arise in various
contexts \cite{noscale}. In the framework of superstring theories
an instance is provided by weakly coupled $E_8\times E_8$ heterotic string
theory compactified on a Calabi-Yau manifold: if the overall modulus field,
whose scalar component represents the size of the compactified
space, dominates the SUSY breaking, as it is the case when this is triggered by gaugino
condensation, the SSB scalar masses, as well as the
trilinear scalar couplings, vanish \cite{heterotic}.
Analogous high energy structures have been shown to appear also in heterotic $M$-theory \cite{Mtheory}.
Furthermore, a no-scale SSB pattern appears with gaugino mediated SUSY breaking in extra dimensional brane models \cite{gauginomediated}. The visible and hidden sectors live on different 3-branes, and SUSY breaking is communicated through the MSSM gauge superfields propagating in the bulk. Since scalars are separated from the SUSY breaking brane, they get negligible soft breaking masses at $M_{\rm bc}$.\\
In models with universal gaugino masses, the no-scale SSB scalar mass boundary
condition $m_0=0$ is not compatible with low energy phenomenology, if the
input scale $M_{\rm bc}$ is taken to be the scale of Grand Unification
$M_{\sss\rm GUT}\simeq 2\times 10^{16}\ {\rm GeV}$ \cite{boundm0}.
If $m_0=0$ at the scale where SM gauge couplings unify, in fact, the mass of the lightest stau turns out to be always {\em smaller} than the mass of the lightest neutralino, regardless of the value of the trilinear scalar coupling $A_0$. Since the LSP is required to be electrically and color neutral, as indicated by stringent cosmological bounds  \cite{chargedlsp}, setting $m_0=0$ at $M_{\sss\rm GUT}$ is ruled out \cite{boundm0}. Nevertheless, as pointed out in \cite{minimalgaugino,boundm0}, if the SSB input scale is {\em larger} than
$M_{\sss\rm GUT}$, RG evolution, driven by GUT-dependent interactions, can
shift $m_0$ from zero to some non-vanishing (and possibly non-universal) value at $M_{\sss\rm GUT}$,
rendering the model compatible with the above mentioned
constraint.\\
The effects of GUT interactions on the weak-scale phenomenology of supersymmetric models have been since long investigated \cite{GUTrunningeffects,su5}. In particular, the case of vanishing scalar masses at a high energy scale has been studied in a {\em minimal} gaugino mediation setting in \cite{minimalgaugino}, where the parameter space was taken to be ($M_{\rm bc}$,$M_{1/2}$) and $B_0=0$ at $M_{\rm bc}$, hence fixing $\tan\beta$ by the radiative electroweak symmetry breaking (REWSB) conditions. In \cite{reach,highenergy}, instead, in order to fulfill Yukawa coupling unification (YU), the $B_0=0$ assumption was relaxed, and the parameter space was taken to be ($\tan\beta$,$M_{1/2}$) at fixed $M_{\rm bc}$. In this paper we resort to this second possibility, focusing on $SU(5)$
and general $SO(10)$ GUTs \cite{highenergy}. In both cases, consistently with the GUT structure, we require $b$-$\tau$ Yukawa coupling unification (YU). We resort to a {\em top-down} approach \cite{btaunonuniversal}, imposing exact unification at $M_{\sss\rm GUT}$ and setting
$h_b(M_{\sss\rm GUT})=h_\tau(M_{\sss\rm GUT})$ in order to obtain at the weak scale the
central experimental value of $m_\tau(M_Z)$. We then compute the
$b$-quark mass at $M_Z$ including SUSY corrections
\cite{precisionMSSM}. We consider the model {\em compatible} with $b$-$\tau$ YU if the calculated value of $m_b^{\rm corr}(M_Z)$ lies within the 95\% C.L. range $m_b(M_Z)=2.83\pm0.22$ \cite{mbmass}.\\
At the scale $M_{\rm bc}$ we set to zero both the soft breaking
scalar masses and the trilinear scalar couplings, while allowing
non-vanishing Higgs mixing masses $B$, supersymmetric higgsino
mixing term $\mu$ and gaugino masses. The latter are supposed to
be universal at $M_{\sss\rm GUT}$. $B$ and $\mu$ are traded for $M_Z$ and
$\tan\beta$ through REWSB. Therefore we are left with only two parameters,
$M_{1/2}$, the universal value of gaugino masses at the GUT scale,
and $\tan\beta$. We start with a trial value for $M_{\sss\rm GUT}$,
$\alpha_{\sss\rm GUT}$ and for the top and $b$-$\tau$ Yukawas, and run them up
to $M_{\rm bc}$, according to the particular chosen GUT model. Evolving the SSB masses and trilinear
 couplings from $M_{\rm bc}$ to $M_{\sss\rm GUT}$, again depending on the GUT, we obtain
a first approximation of the SSB structure of the model at
$M_{\sss\rm GUT}$. Further evolution down to the weak scale, performed
with the ISAJET package \cite{isajet}, allows to adjust the value
of the Yukawas and to find, consistently with the low energy SUSY
effective threshold and with the SM gauge coupling running, the
values of $M_{\sss\rm GUT}$ and $\alpha_{\sss\rm GUT}$. The whole loop is then
repeated until convergence is found. In practice we find that
three loops are sufficient to stabilize, for consistency, the SSB
pattern as well as the Yukawas and the GUT
scale and coupling.\\
At the low energy scale, besides successful $b$-$\tau$ YU, we
require the fulfillment of the known phenomenological constraints.
We find that the most stringent bound comes from the inclusive branching
ratio $BR(b\rightarrow s\gamma)$, which we require to lie in the
following range \cite{constraints}:
\begin{equation}
2.16\times 10^{-4}<BR(b\rightarrow s\gamma)<4.34\times10^{-4}.\label{bsg}
\end{equation}
Finally, we compute, using the latest version of the {\tt
DarkSUSY} code \cite{darksusy}, the resulting neutralino relic
density $\Omega_\chi h^2$ and the spin independent
neutralino-proton cross section $\sigma^{\rm SI}_{\chi p}$. The
recent data from the WMAP survey, combined in a global fit with
other CMB, large scale structure and Ly$\alpha$ data, constrain
the relic density to be
\begin{equation}
\displaystyle \Omega_{\sss\rm CDM}\ h^2\ =\
0.1126^{+0.00805}_{-0.00905}. \label{omegarange}
\end{equation}
As the lower limit can be evaded under the hypothesis of the
existence of another cold dark matter component besides
neutralinos, we take here as a constraint only the 95\% C.L. upper bound
$\Omega_{\chi}h^2\lesssim0.1287$.

\subsection{The no-scale $SU(5)$ case}
\begin{figure*}
\begin{center}
\begin{tabular}{c}
\includegraphics[scale=0.55]{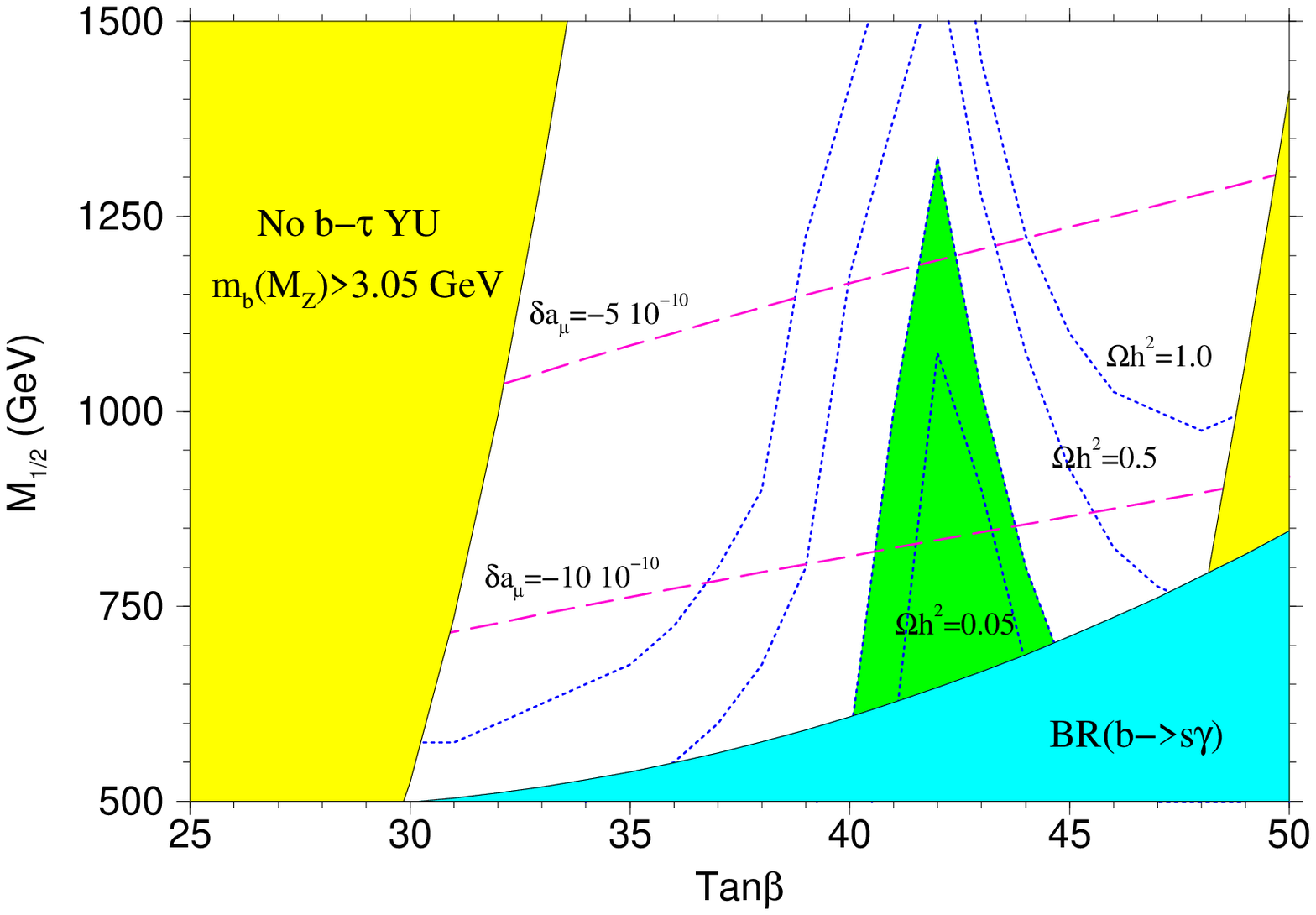}\\
\includegraphics[scale=0.55]{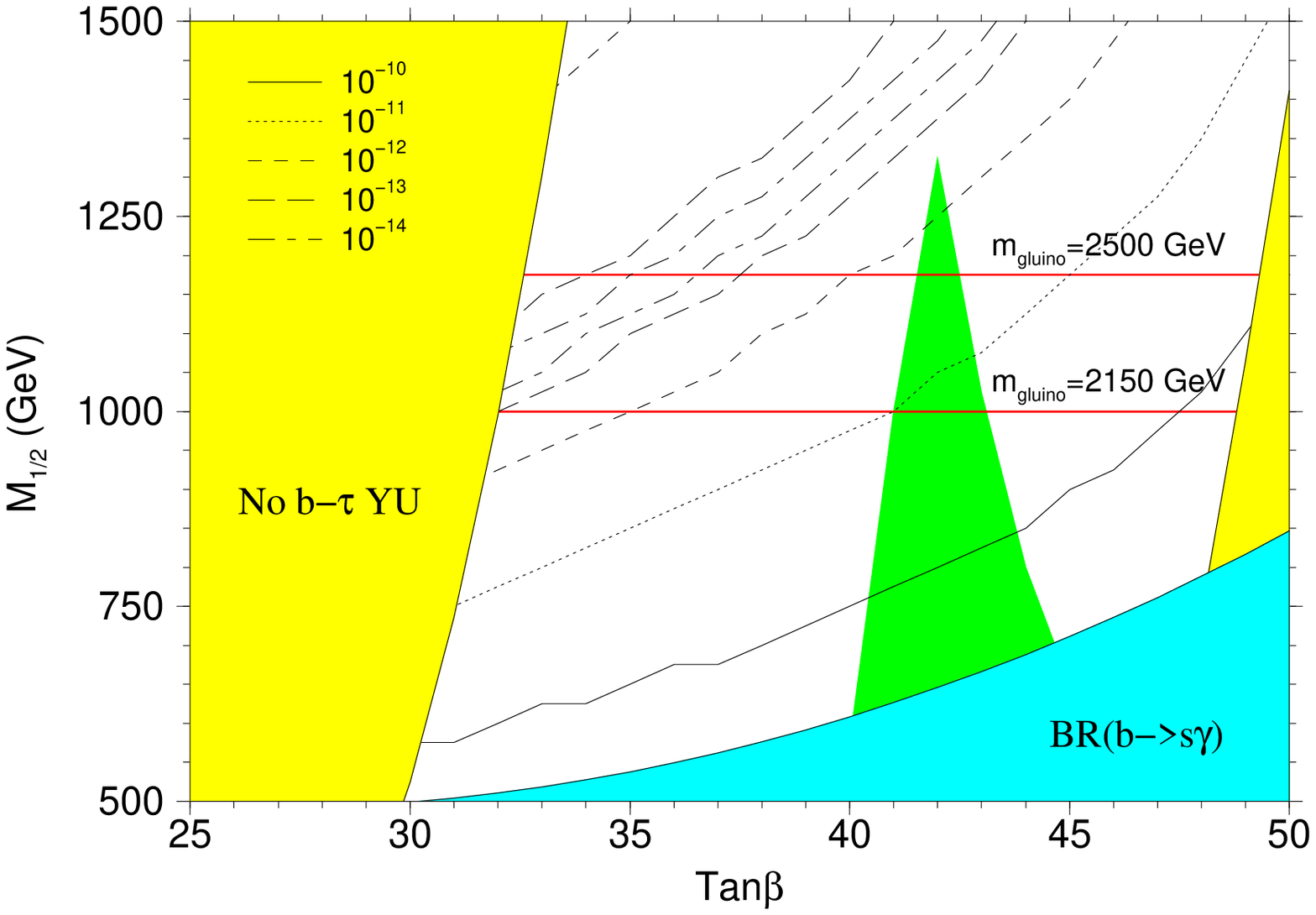}
\end{tabular}
\end{center}
\caption{\label{su5detect}\label{su5cosmo} Isolevel curves for the neutralino relic density and muon anomalous magnetic moment ({\em upper panel}), and for direct WIMP detection and accelerator searches at LHC ({\em lower panel}) in the case of $SU(5)$ GUT.}
\end{figure*}
\begin{figure*}
\begin{tabular}{c c}
\includegraphics[scale=0.44]{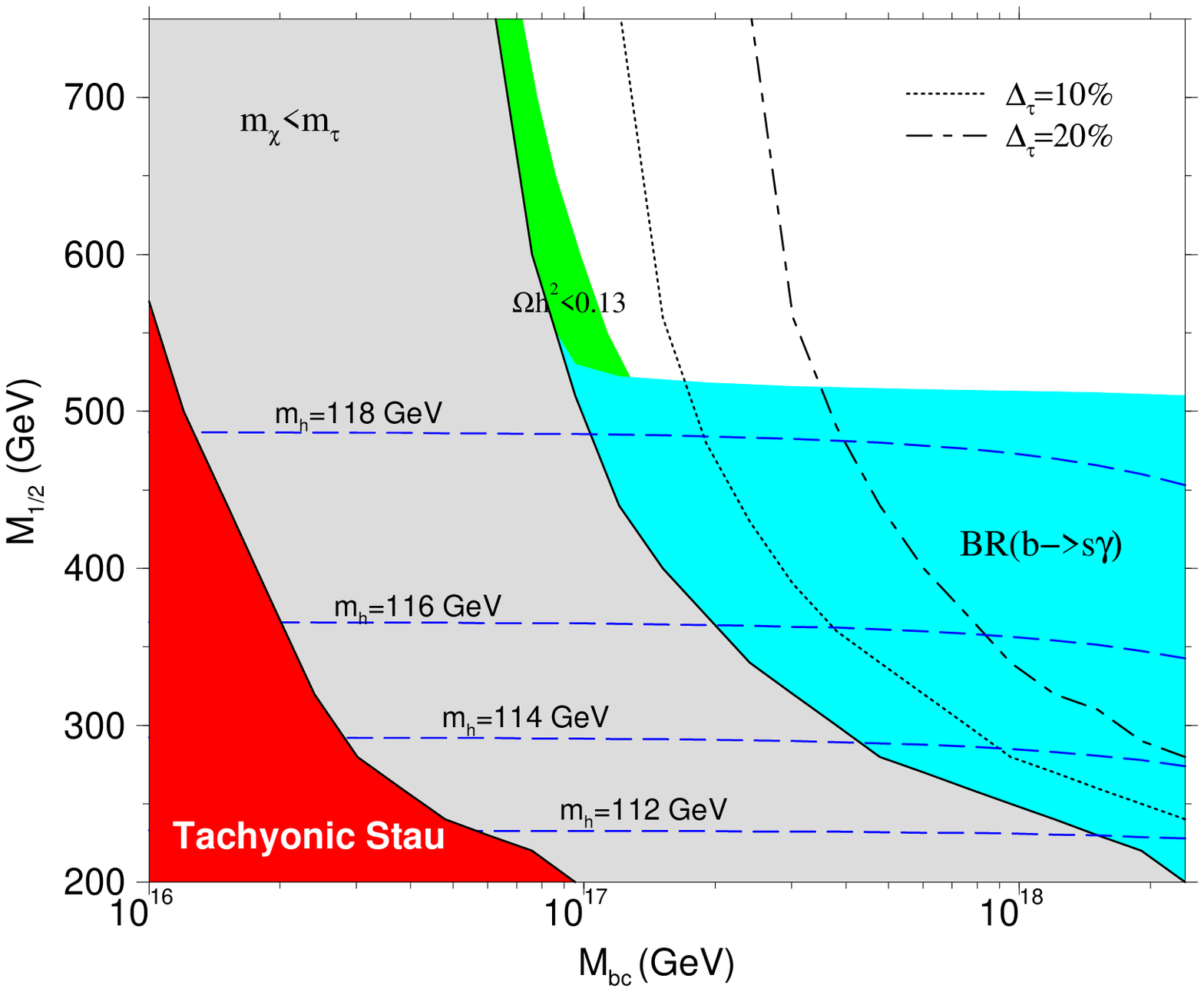}&
\includegraphics[scale=0.44]{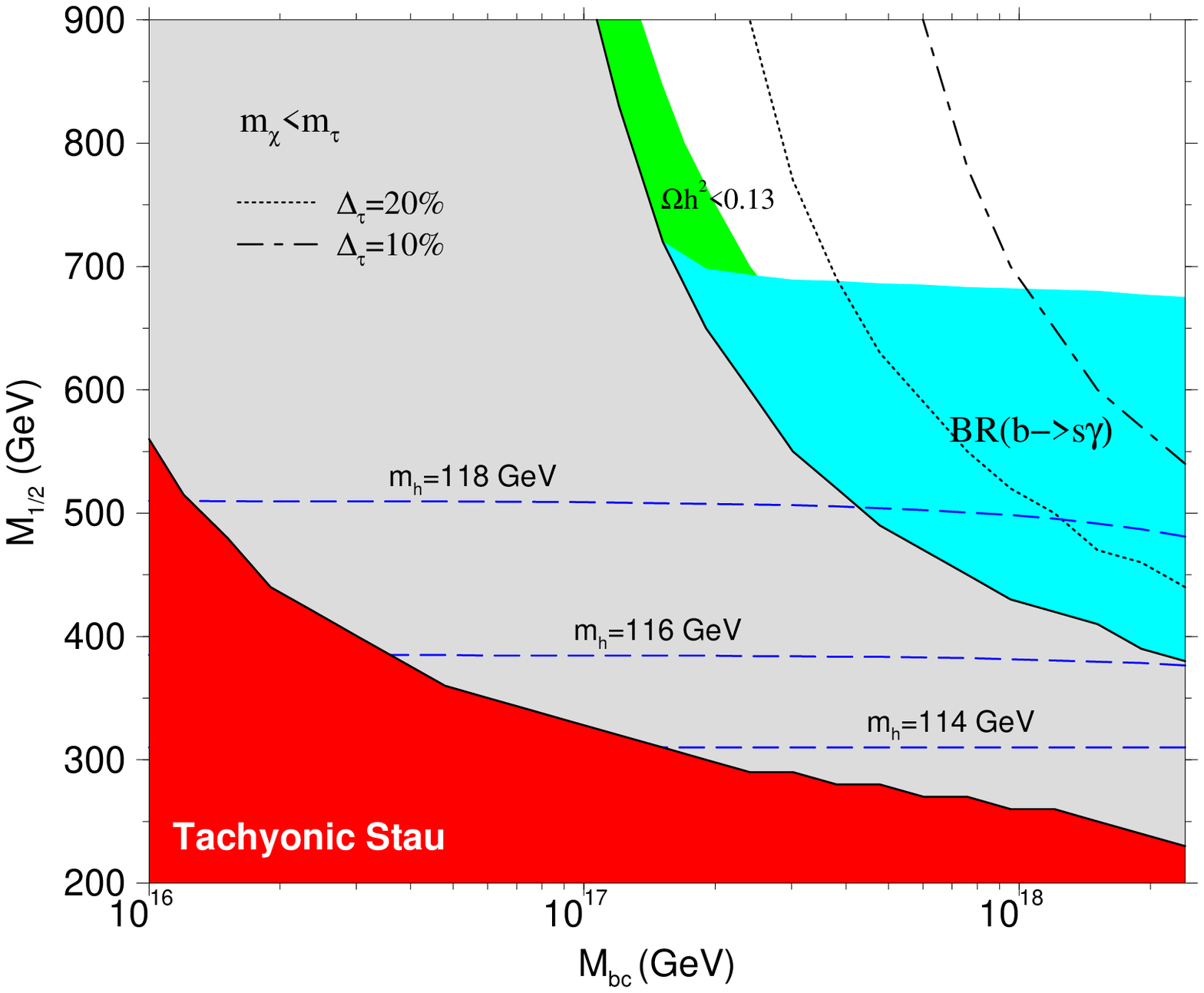}
\end{tabular}
\caption{\label{scale}Allowed parameter space at various boundary condition input scale $M_{\rm bc}$,  in the case of $SU(5)$ GUT, at $\tan\beta=35.0$  ({\em left}) $\tan\beta=47.0$  ({\em right}).}
\end{figure*}
\begin{figure*}
\begin{center}
\begin{tabular}{cc}
\includegraphics[scale=0.40]{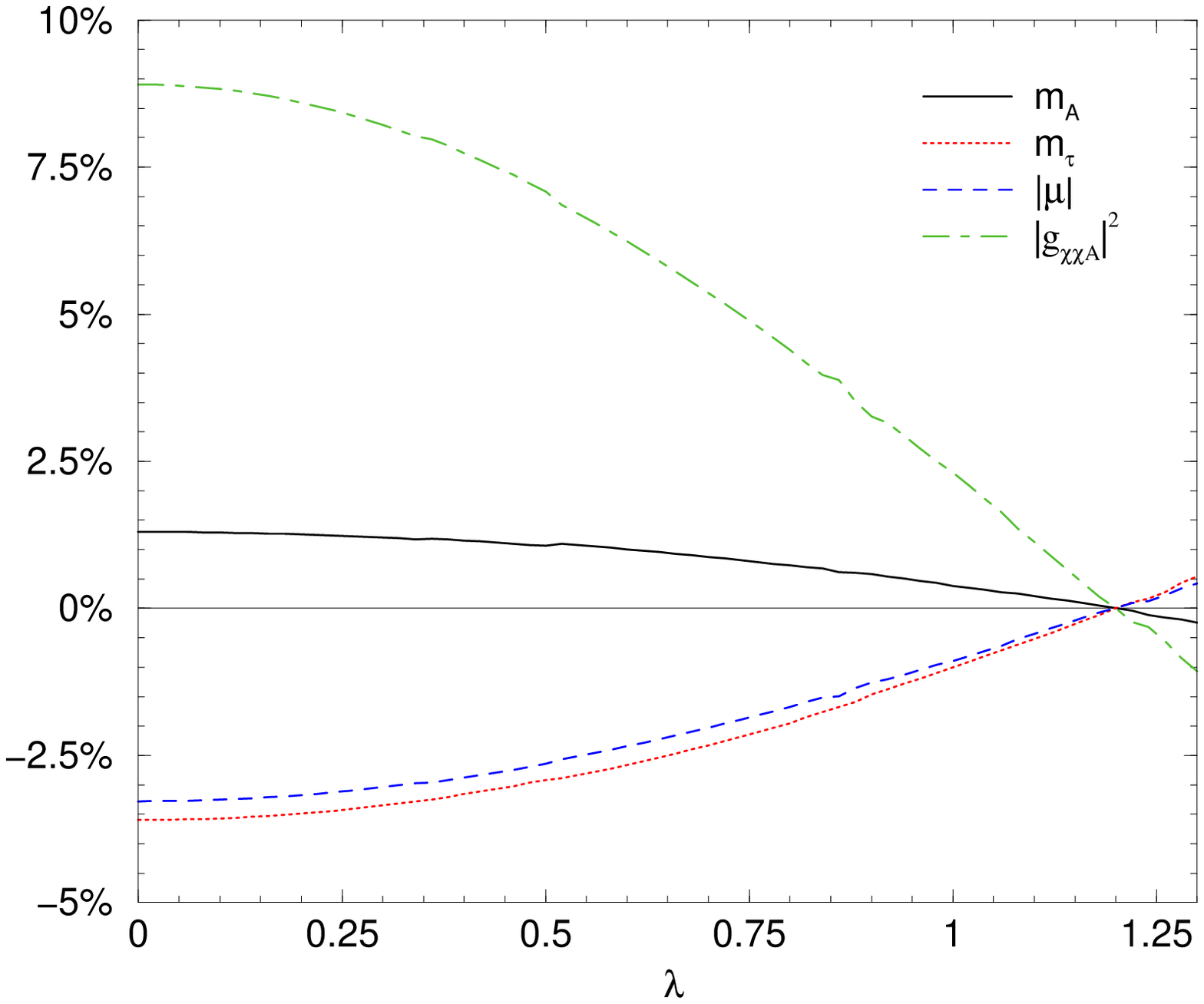}&
\includegraphics[scale=0.40]{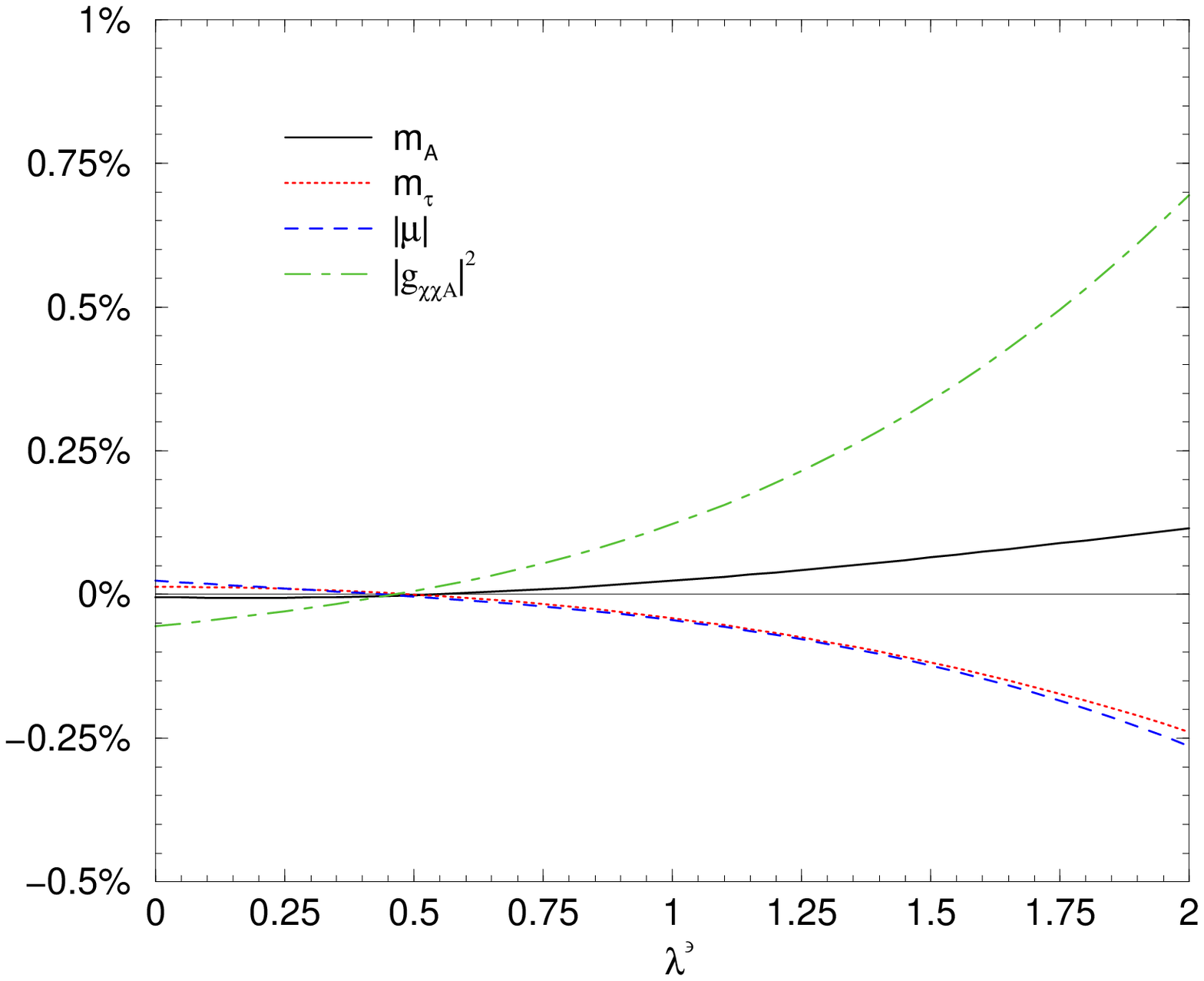}
\end{tabular}
\end{center}
\caption{\label{su5lambda} ({\em Left}): the percent splitting, from the reference value at $\lambda=1.2$, of $m_A$, $m_{\tilde\tau_1}$, $|\mu|$ and $|g_{\chi\chi A}|^2$ at various $\lambda$'s, and at fixed $\lambda^\prime=0.5$. ({\em Right}): the same at fixed $\lambda=1.2$ and at various $\lambda^\prime$; the reference value is $\lambda^\prime=0.5$. In both frames $\tan\beta=43.0$, $M_{1/2}=1000\ {\rm GeV}$ and $M_{\rm bc}=M_{\rm Pl}$.}
\end{figure*}

In the minimal $SU(5)$ GUT model \cite{su5} the matter content of the MSSM
is collected into a $\overline{\bf 5}$ ($\hat{D}^c$, $\hat{L}$)
and a ${\bf 10}$ \mbox{($\hat{Q}$, $\hat{U}^c$, $\hat{E}^c$)}
supermultiplets. The Higgs sector contains three supermultiplets:
$\hat{\Sigma}({\bf 24})$, responsible for the $SU(5)$ breaking to
the SM gauge group, $\hat{H}_1({\bf 5})$ and $\hat{H}_2({\bf 5})$
containing the MSSM Higgs doublet superfields $\hat{H}_d$ and
$\hat{H}_u$. The superpotential reads:
\begin{eqnarray}\label{superpotential}
W_\Sigma & = &\mu_\Sigma{\rm
Tr}\hat{\Sigma}^2+\frac{1}{6}\lambda^\prime{\rm
Tr}\hat{\Sigma}^3+\lambda\hat{H}_1\hat{\Sigma}\hat{H}_2+\mu_{H}\hat{H}_1\hat{H}_2\\
 & & +\frac{1}{4}h_t({\bf 10})({\bf 10})\hat{H}_2+\sqrt{2}h_b({\bf 10})(\overline{\bf 5})\hat{H}_1.
\end{eqnarray}
The boundary conditions at the  scale $M_{\rm bc}$ for scalar masses and trilinear couplings are taken to be \cite{reach}:
\begin{eqnarray*}
&m_{10}=m_5=m_{H_1}=m_{H_2}=m_{\Sigma}=0,&\\
&A_t=A_b=A_\lambda=A_\lambda^\prime=0.&
\end{eqnarray*}
The Higgs couplings $\lambda(M_{\sss\rm GUT})$ and
$\lambda^\prime(M_{\sss\rm GUT})$ which appear in the superpotential
(\ref{superpotential}) are constrained, for the stability of the
RGE evolution, in the range 
\begin{equation}\label{lambdabounds}
|\lambda(M_{\sss\rm GUT})|\lesssim 1.5, \quad |\lambda^\prime(M_{\sss\rm GUT})|\lesssim 3.0~.
\end{equation}
The coupling $\lambda(M_{\sss\rm GUT})$ is related to the mass of the colored
Higgs triplet responsible of rapid proton decay, which in the context of {\em minimal} $SU(5)$ has since long been recognized as a critical issue \cite{oldprotondecay}. It has been recently claimed that the minimal setting outlined above is {\em ruled out} by the current experimental limits on the proton lifetime \cite{protondecay} (see however \cite{protonnodecay}) coming from the SuperKamiokande results \cite{superkamiokande}. Nevertheless, consistent $SU(5)$ models have been recently proposed in ref.~\cite{consistentsu5}. In these next-to-minimal scenarios, suitable structures for the mixed Yukawa couplings $Y_{QQ}$, $Y_{UE}$, $Y_{UE}$ and $Y_{QL}$, determined by the inclusion of dimension five operators, can drastically suppress the proton decay rate, even at large $\tan\beta$, below the current experimental limits. Even though addressing the issue of proton decay goes beyond the scope of the present paper, we stress that the mentioned Yukawa structures do not significantly affect the SUSY spectrum we deal with, and can be consistently neglected for the present purposes, as it is the case for the Yukawa couplings of the two lightest generations. 
To maximize the enhancement of the proton life-time \cite{oldprotondecay}, we nevertheless resort to values of $\lambda(M_{\sss\rm GUT})$ close to the upper bound (\ref{lambdabounds}). We fix $\lambda(M_{\sss\rm GUT})=1.2$ and $\lambda^\prime(M_{\sss\rm GUT})=0.5$, and show at the end of this section that different choices of the two parameters would not significantly affect our results.\\
We start our analysis setting for definiteness $M_{\rm bc}=M_{\rm Pl}=2.4\times 10^{18} \ {\rm GeV}$. Top-down $b$-$\tau$ YU  constrains the sign of $\mu$ to be negative \cite{btaunonuniversal}, and forces $\tan\beta$ to large values $\gtrsim28$. We show in the upper panel of fig.~\ref{su5cosmo} the allowed parameter space in the ($\tan\beta$,$M_{1/2}$) plane and the curves at fixed $\Omega_\chi h^2=0.05,\ 0.13,\ 0.5,\ 1.0$. The yellow regions on the left (right) part of the figures do not fulfill $b$-$\tau$ YU, giving rise to \mbox{$m_b(M_Z)>3.05\ {\rm GeV}$} \mbox{(resp. $m_b(M_Z)<2.61\ {\rm GeV}$)}. Values of $\tan\beta$ higher than 51, besides being excluded by $b$-$\tau$ YU, do not fulfill REWSB. Since the SUSY contributions $\delta m_b\propto(-\tan\beta\cdot f(M_{1/2}))$, with $f(M_{1/2})$ a {\em decreasing} function of $M_{1/2}$, as we increase $M_{1/2}$ the bounds on $m_b(M_Z)$ are saturated at larger values of $\tan\beta$. The light blue lower part of the figure is ruled out by the $BR(b\rightarrow s\gamma)$ constraint (\ref{bsg}). This bound, for $\mu<0$, becomes stronger at higher $\tan\beta$ and lower $M_{1/2}$, thus the shape of the excluded region is easily understood. We also plot isolevel lines for the SUSY contributions to the muon anomalous magnetic moment $\delta a_\mu$. However, due to the current theoretical uncertainties in the evaluation of the SM hadronic contribution, we do not use this quantity as a constraint. Finally, we find that in the whole allowed parameter space $m_{\chi}\simeq0.44M_{1/2}$ to within few percent of accuracy.\\
As regards the cosmologically preferred green region, satisfying (\ref{omegarange}), and the behavior of the $\Omega_\chi$ isolevel curves, we find, in the range $40\lesssim\tan\beta\lesssim45$, that $2m_\chi\approx m_A$, $m_A$ being the $CP$-odd neutral Higgs $A$ mass\footnote{Funnel regions do not occur, instead, in the models of ref.\cite{minimalgaugino}, since the $B_0=0$ condition forces $\tan\beta$ to low values where $2m_\chi\approx m_A$ cannot be fulfilled.}. The line at  $2m_\chi= m_A$ lies at $\tan\beta\approx43$, while at higher (resp. lower) values of $\tan\beta$ \mbox{$2m_\chi>(<)m_A$}. 
The overall bounds on the parameter space of the model are $40\lesssim\tan\beta\lesssim45$ and $590\ {\rm GeV}\lesssim M_{1/2}\lesssim 1330\ {\rm GeV}$, which translates into a bound for the lightest neutralino mass of $250\ {\rm GeV}\lesssim m_{\chi}\lesssim 585\ {\rm GeV}$. We also find that for $M_{\rm bc}=M_{\rm Pl}$ the allowed parameter space {\em excludes} coannihilation effects with the next-to-lightest SUSY particle, the lightest stau, always lying more than $\approx25\%$ above the LSP mass.\\ 
The lower panel of fig.~\ref{su5detect} summarizes the detection perspectives of the model under scrutiny, both at direct WIMP detection experiments \cite{directdetection} (spin-independent $\sigma^{\rm SI}_{\chi p}$ isolevel curves; the values are in {\em pb}) and at CERN LHC \cite{reach}. Detection rates are beyond reach of {\em stage 2} detectors (CDMS2, EDELWEISS2, ZEPLIN2), while the low $M_{1/2}$ and large $\tan\beta$ part of the cosmologically preferred region could be within reach of the so-called {\em stage 3} detectors (GENIUS, ZEPLIN4, CRYOARRAY). We notice in the upper-left part of the figure a dip in $\sigma^{\rm SI}_{\chi p}$, due to cancellations among terms stemming from up and down type quarks interactions, the largest contributions then coming from $t$-channel Higgs boson exchanges \cite{directellis}. As regards LHC searches, we expect detectability, in the present framework, following the results of \cite{reach}, for $m_{\tilde g}\lesssim2150\ {\rm GeV}$ ($2500\ {\rm GeV}$) at an integrated luminosity of 10 (100) ${\rm fb}^{-1}$. We show in the lower panel of fig.~\ref{su5detect} with red solid lines the curves at $m_{\tilde g}=2150,\ 2500\ {\rm GeV}$: in both cases, for $\tan\beta\approx43$ and $M_{1/2}\gtrsim1160\ {\rm GeV}$, i.e. $m_{\chi}\gtrsim 500\ {\rm GeV}$, a slice of parameter space compatible with cosmological requirements may be beyond the reach of CERN LHC.\\
In fig.~\ref{scale} we show, at two different values of $\tan\beta=35.0$ and 47.0, respectively representative  of the lower and upper part of the $b$-$\tau$ YU allowed range, the constraints on $M_{1/2}$ at various values of the no-scale boundary condition scale $M_{\rm bc}$. The phenomenologically allowed region is in the upper right corner of the plots, where the isolevel curves for the NLSP mass splitting $\displaystyle\Delta_{\tilde\tau}\equiv \frac{m_{\tilde\tau_1}-m_\chi}{m_\chi}$ are also depicted at $\Delta_{\tilde\tau}=0\%,\ 10\%,\ 20\%$. The light blue region is excluded by the $BR(b\rightarrow s\gamma)$ constraint (\ref{bsg}), while in the gray region $m_{\chi}<m_{\tilde\tau_1}$. Finally, the red region in the lower left corner has $m^2_{\tilde\tau_1}<0$. The lowest possible value for $M_{\rm bc}$ is around $5\times10^{16}\ {\rm GeV}$, which is reached at the lowest $\tan\beta$ compatible with $b$-$\tau$ YU, and, at a given $M_{1/2}$, is always above the corresponding $M_{\sss\rm GUT}$. For completeness, we also include the curves at fixed values of the lightest neutral Higgs $m_h$, for which the LEP2 bound gives $m_h\gtrsim114.1\ {\rm GeV}$ \cite{lephiggs}. We notice that the bound on the Higgs mass, which in the MSSM at large $\tan\beta$ is even milder than the mentioned value, is always weaker than the other considered phenomenological constraints.\\
As regards the cosmologically viable parameter space, the effect of reducing $M_{\rm bc}$ translates in the appearance of coannihilation regions: we recall that in order to effectively reduce the neutralino relic density, $\Delta_{\tilde\tau}$ must be less than $\approx10\%$, depending on the absolute value of $m_\chi$, and this is the case for 
\begin{equation}\label{mbcrange}
5\times10^{16}\ {\rm GeV}\lesssim M_{\rm bc}\lesssim 3\times10^{17}\ {\rm GeV}.
\end{equation} 
We shade in green in fig.~\ref{scale} the actual parameter space regions where $\Omega h^2<0.13$. 
Furthermore, the $A$-pole condition $2m_{\chi}\simeq m_A$ is still fulfilled for $M_{\rm bc}<M_{\rm Pl}$, always at $\tan\beta$ close to 43. Therefore, in the range (\ref{mbcrange}) an interplay between neutralino-stau coannihilations and direct $A$-pole annihilations can significantly enlarge the cosmologically preferred parameter space of the models. Outside the funnel region, however, a certain fine-tuning between $\tan\beta$ and $M_{\rm bc}$ is needed in order to enter the coannihilation region (see again fig.~\ref{scale}). Finally, we notice that the isolevel stau mass splitting curves are {\em steeper} for lower $\tan\beta$. Therefore, at low $\tan\beta$, coannihilations suppress $\Omega_\chi h^2$ to viable values in a {\em wider range} of $M_{1/2}$ with respect to the high $\tan\beta$ regime: one can clearly understand this statement imaging vertical lines (i.e. lines at fixed $M_{\rm bc}$) intersecting the green regions in the left ($\tan\beta=35.0$) and right ($\tan\beta=47.0$) panels of fig.~\ref{scale}.\\
We now turn to the question of the sensitivity of the overall outlined scenario on the particular chosen values of the couplings $\lambda(M_{\sss\rm GUT})$ and $\lambda^\prime(M_{\sss\rm GUT})$. We focus on the variations induced in $m_A$ and in $m_{\tilde\tau_1}$, which are directly responsible for the relic density suppression, as well as in the neutralino-neutralino-Higgs couplings, on which both the neutralino relic density $\Omega_\chi h^2$ and the scattering cross-section $\sigma^{\rm SI}_{\chi p}$ depend\footnote{I thank the Referee for having drawn my attention to this point.}. These couplings depend, in their turn, on the absolute value of $\mu$, which we therefore also take into account. For definiteness, we choose the representative coupling $|g_{\chi\chi A}|^2$, as defined in ref.~\cite{gunionhaber}; we checked that the behavior of the other neutralino-neutralino-Higgs couplings is very similar. The quantities we study are the percent differences between the values at a given $\lambda$ (resp. $\lambda^\prime$) and those at the particular one we picked, namely  $\lambda=1.2$ (resp. $\lambda^\prime=0.5$).\\ 
In the left panel of fig.~\ref{su5lambda}, where the dependence on $\lambda$ is investigated, we find, as expected \cite{su5}, that the variations of $m_{A}$ are rather mild (at most $\sim1\%$); instead, the $\mu$ parameter and the mass of the lightest stau are more sensitive to $\lambda$: in the range of perturbative $\lambda$'s which we study, they vary within a $\sim5\%$ range. The resulting variations of $|g_{\chi\chi A}|^2\propto1/|\mu|^2$ can be as large as 10\%. Though $m_A$ varies rather slowly with $\lambda$, we find that in the range of large $\tan\beta$ where $b$-$\tau$ YU is fulfilled, the effects on the relic density and on $\sigma^{\rm SI}_{\chi p}$ driven by the quantity \mbox{$\Delta_A\equiv m_A-2m_\chi$} largely {\em dominate} on the corrections due to the variations of the neutralino-Higgs couplings. The reason is that in this range of $\tan\beta$ both  nucleon scattering and neutralino annihilations are dominated by the heavy Higgses ($A$ and $H$) resonances. Therefore, if $m_A\gtrsim2m_\chi$ (i.e. $\tan\beta\lesssim43$) $\Omega_\chi h^2$ (resp. $\sigma^{\rm SI}_{\chi p}$) {\em decreases} (resp. {\em increases}) with increasing $\lambda$, because of the induced reduction on $\Delta_A$ (see fig.~\ref{su5lambda}). The opposite mechanism takes place at $\tan\beta\gtrsim43$.\\
To summarize the effects of varying $\lambda(M_{\sss\rm GUT})$, we find a superposition of two phenomena. On the one side, increasing $\lambda$ decreases $m_A$, and therefore {\em shifts} the funnel depicted in fig.~\ref{su5detect}. 
On the other side, increasing $\lambda$ suppresses the neutralino-neutralino-Higgs couplings, and therefore, at fixed $\Delta_A$, slightly reduces $\sigma^{\rm SI}_{\chi p}$ and increases $\Omega_\chi h^2$.   We find that at a given ($\tan\beta,M_{1/2}$) point this second mechanism, in the present large $\tan\beta$ range, is subdominant with respect to the first one: the kinematic condition $\Delta_A\simeq0$ is more critical than the size of the neutralino-Higgs couplings. To quantify the {\em overall size} of these variations, we find that at $\lambda(M_{\sss\rm GUT})=0$ the position of the funnel is shifted to the right, with respect to the plotted case $\lambda(M_{\sss\rm GUT})=1.2$, by $\approx0.3$ units in $\tan\beta$, while the width of the funnel\footnote{By {\em width} we mean the $\tan\beta$ range at fixed $M_{1/2}$ where $\Omega h^2<0.13$.} is enlarged by a factor $\approx 2\%$. This last effect is purely due to the variations of the neutralino-Higgs couplings with $\lambda(M_{\sss\rm GUT})$. The corrections on $\sigma^{\rm SI}_{\chi p}$ are also very small, and not appreciable in a plot like the lower panel of fig.~\ref{su5detect}.\\
As far as the variations of the lightest stau mass with $\lambda$ are concerned, though irrelevant at $M_{\rm bc}=M_{\rm Pl}$, they could in principle affect the position of the coannihilation regions depicted in fig.~\ref{scale}. Nevertheless, as $M_{\rm bc}$ is reduced, the size of the $\lambda$-dependent effects on the spectrum is suppressed, and, as a consequence, also in this case the low $\lambda(M_{\sss\rm GUT})$ scenario would closely resemble the one studied here. The only consequence of taking a lower value for $\lambda$ is a shift of the coannihilation regions, and of the stau mass isolevel curves in fig.~\ref{scale} towards larger $M_{\rm bc}$ values.\\
Finally, as regards the variations induced by $\lambda^\prime(M_{\sss\rm GUT})$, we find that the effects are always negligible, as illustrated in the right panel of fig.~\ref{su5lambda} (notice that the scale in the plot is reduced by a factor 10 with respect to the right panel).  

\subsection{The no-scale general $SO(10)$ case}
\begin{figure*}
\begin{tabular}{c c c}
\includegraphics[width=8.5cm,height=6.7cm]{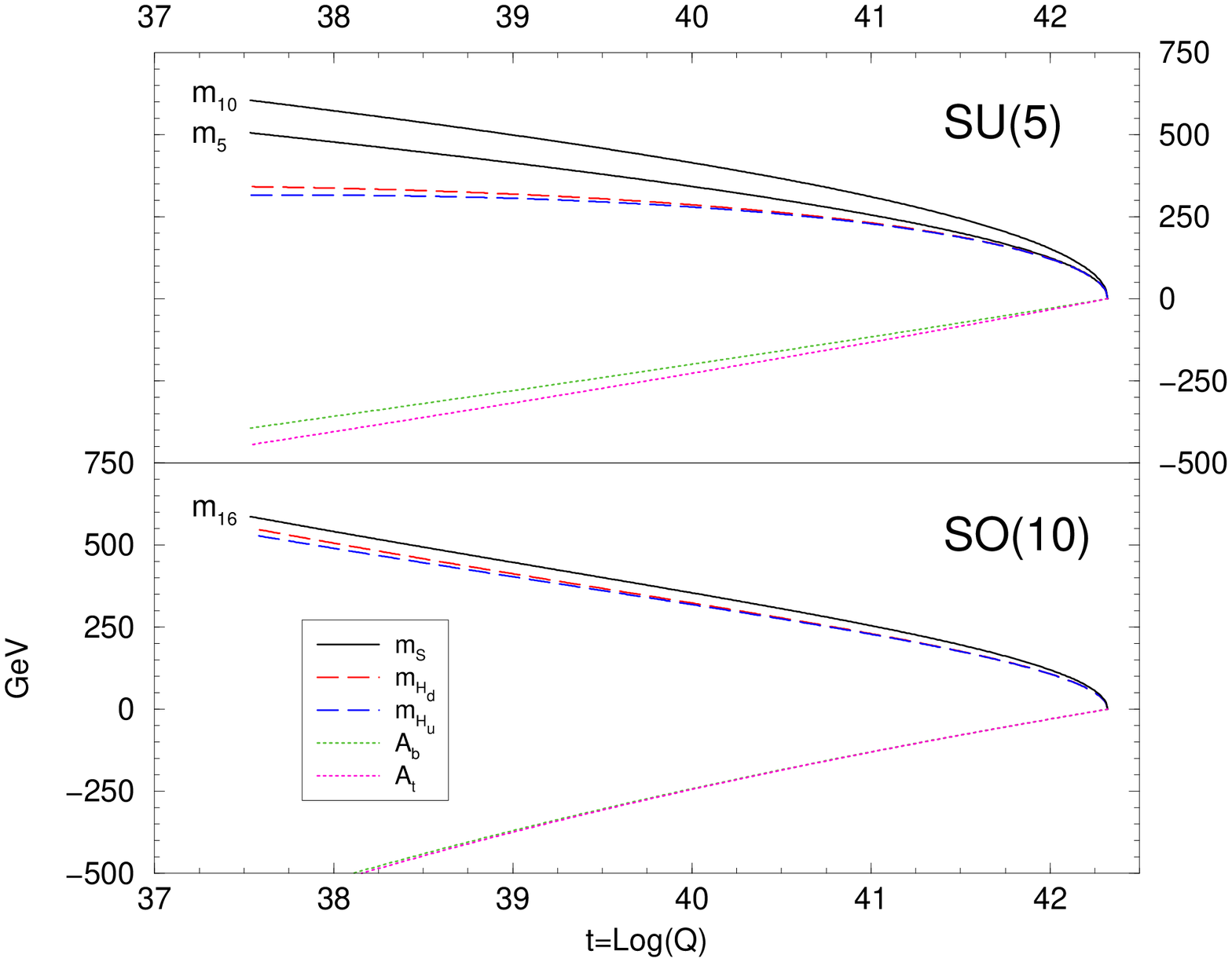}& &
\includegraphics[width=5cm,height=6.5cm]{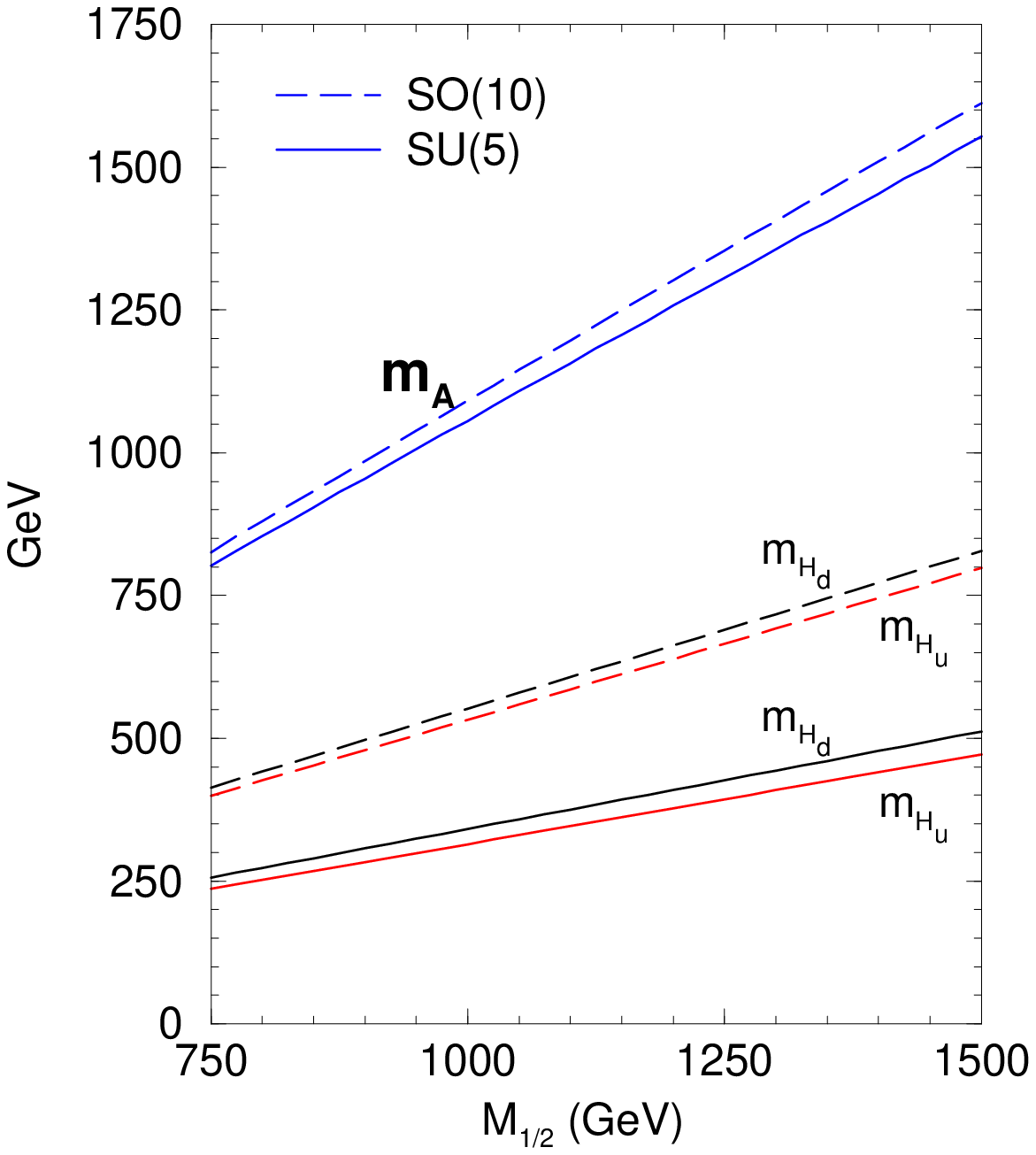}
\end{tabular}
\caption{\label{running}  ({\em Left}): the running of the soft scalar masses and trilinear scalar couplings in $SU(5)$ (upper panel) and $SO(10)$ (lower panel) GUTs for $M_{1/2}=1\ {\rm TeV}$. ({\em Right}): the four lower lines represent the $m_{H_u}$ (black) and $m_{H_d}$ (red) soft scalar masses at the GUT scale as a function of $M_{1/2}$; the two upper lines are the resulting values for $m_A$ at the low energy scale. Solid lines indicate the $SU(5)$ case, while dashed lines $SO(10)$. In both figures $M_{\rm bc}=M_{\rm Pl}$ and $\tan\beta=35.0$.}
\end{figure*}
In minimal SUSY $SO(10)$ GUT the matter superfields of the MSSM plus an
additional gauge singlet right handed neutrino are collected in
a ${\bf 16}$ supermultiplet, while the Higgs superfields belong
to a ${\bf 10}$. 
A top down-like approach to YU like the one we propose here would lead, in the present framework of {\em universal} gaugino masses\footnote{See \cite{nonuniversal} and references therein for the related case of $SO(10)$ with non-universal gaugino masses.}, to a large fine-tuning, at fixed $\tan\beta$, between $M_{1/2}$ and the mandatory non-vanishing $D$-term contribution $M_D^2$ in order to achieve complete YU. We therefore resort to a more {\em general} $SO(10)$ setting \cite{highenergy}, where the two MSSM Higgs doublets live in two different fundamental representations of the GUT gauge group, and fix $M_D^2=0$. In this case the superpotential reads
\begin{equation}
W_{\rm gen}=f_t({\bf 16})({\bf 16})\hat{H}_2+f_b({\bf 16})({\bf
16})\hat{H}_1
\end{equation}
and only $b$-$\tau$ YU is required. The RGE's depend in general on
the Higgs multiplets and on the the number of matter generations,
namely
\begin{eqnarray}
\frac{{\rm d}g}{{\rm d}t}& = & \frac{g^3}{16\pi^2}(S-24)\\
\frac{{\rm d}M_{1/2}}{{\rm d}t}& = & \frac{2}{16\pi^2}(S-24)g^2M_{1/2},
\end{eqnarray}
where $S$ is the sum of the Dynkin indices of the chiral
superfields of the model, $g$ is the $SO(10)$ unified gauge coupling and $M_{1/2}$ is the common gaugino mass above the unification scale. In the case of two Higgs and 3
generations, $S=8$.\\
\begin{figure*}
\begin{center}
\begin{tabular}{c}
\includegraphics[scale=0.55]{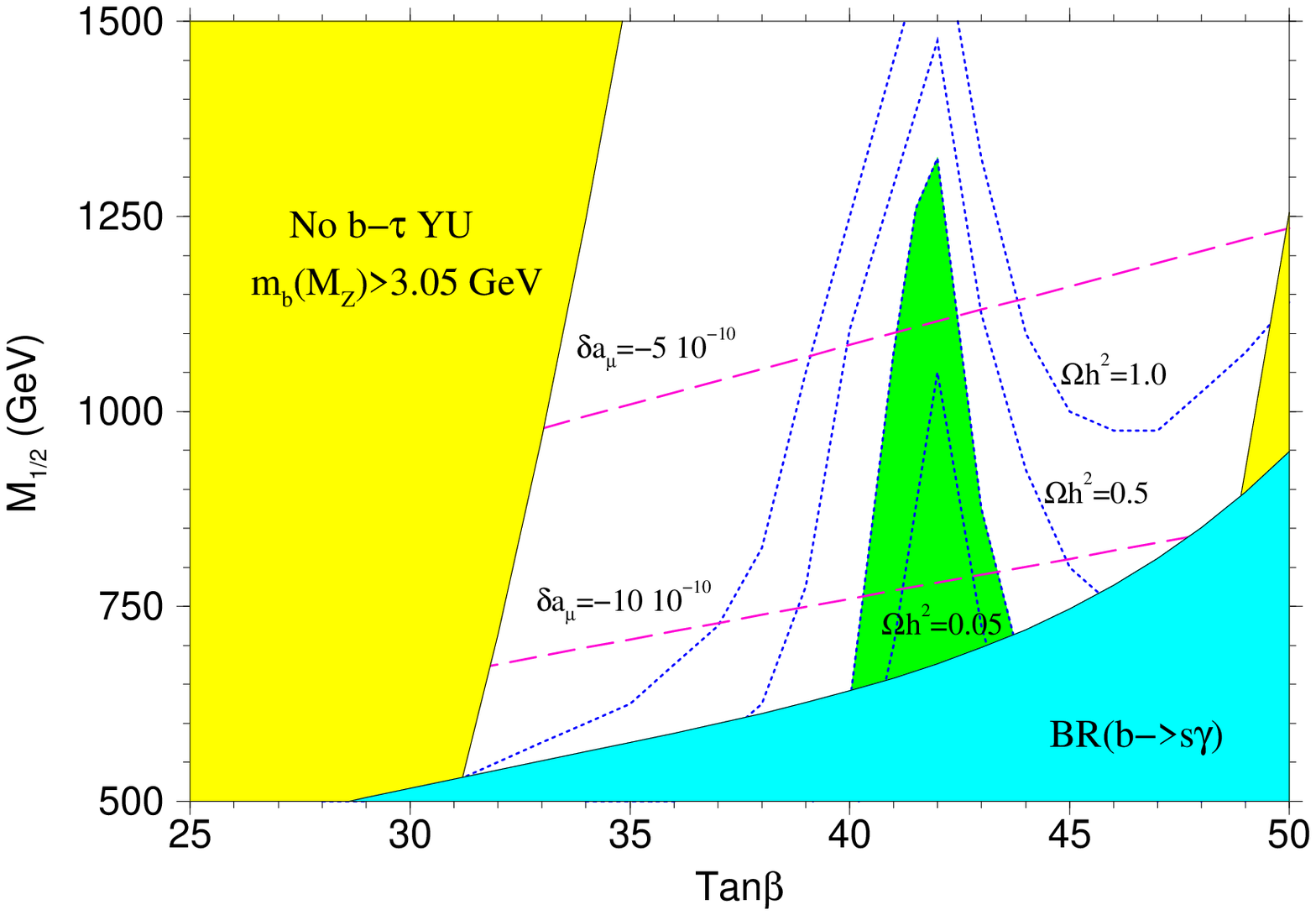}\\
\includegraphics[scale=0.55]{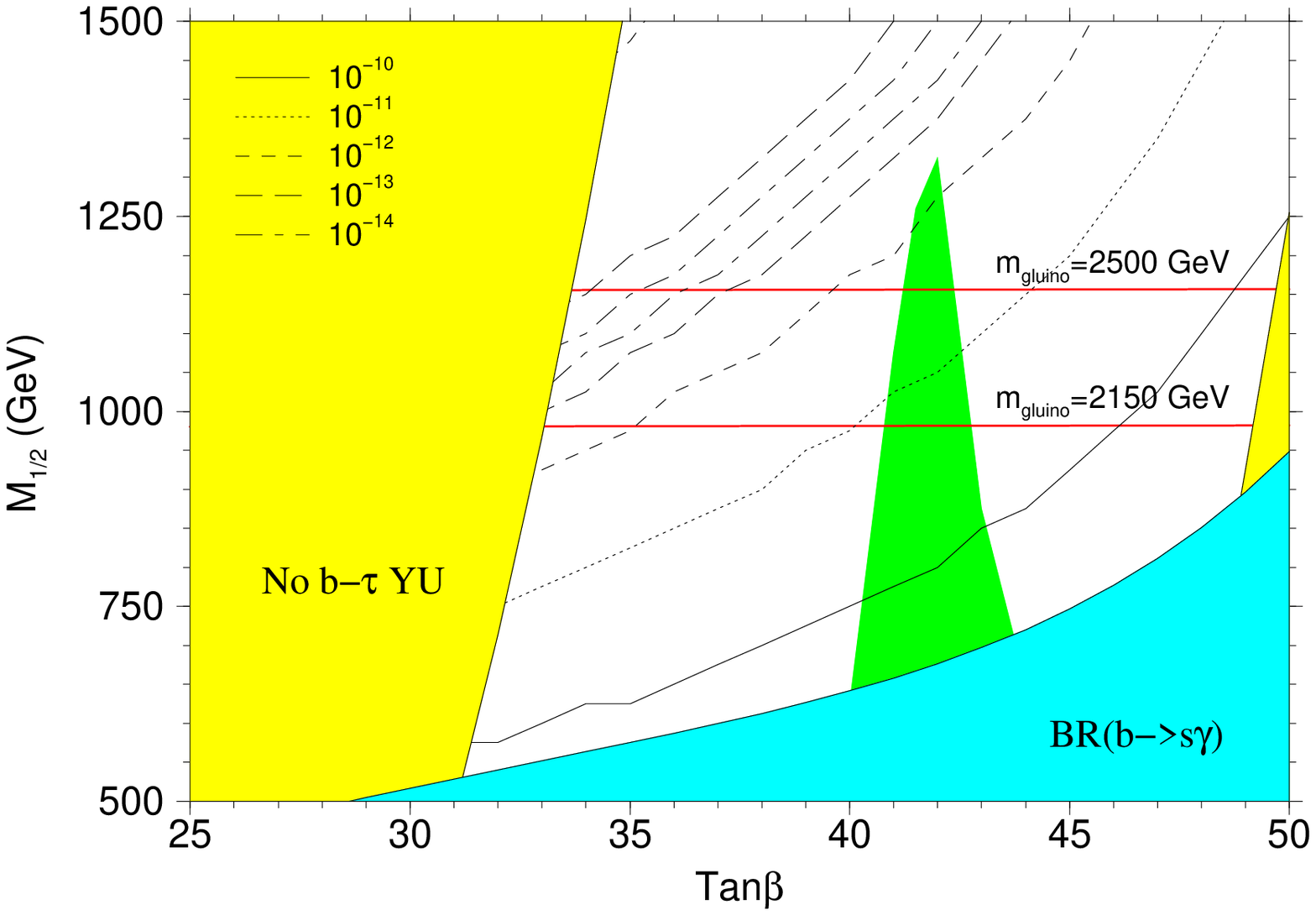}
\end{tabular}
\end{center}
\caption{\label{so10detect}\label{so10cosmo} Isolevel curves for the neutralino relic density and muon anomalous magnetic moment ({\em upper panel}), and for direct WIMP detection and accelerator searches at LHC ({\em lower panel}) in the case of {\em general} $SO(10)$ GUT.}
\end{figure*}
\noindent We begin comparing, in the left panel of fig.~\ref{running}, the effects of the GUT running between $M_{\rm bc}$ (set to $M_{\rm Pl}$ for definiteness) and $M_{\sss\rm GUT}$ for $SU(5)$ and $SO(10)$ GUTs. First, notice that the overall scale at which the soft scalar mass are driven by RG running is comparable in both cases, while the trilinear couplings are evolved towards higher negative values in the $SO(10)$ case. A further remarkable feature is that in $SU(5)$ the Higgs soft SUSY breaking masses are driven to significantly lower values than in  $SO(10)$. Last, notice the larger departure from universality which takes place in the scalar sector of $SU(5)$. Being $m_{10}(M_{\sss\rm GUT})\simeq m_{16}(M_{\sss\rm GUT})$, this translates into a lower soft scalar mass pattern for $SU(5)$.\\
In the right part of  fig.~\ref{running} we investigate the effects induced in the Higgs sector at the low energy scale by the different running, between $M_{\rm bc}$ and  $M_{\sss\rm GUT}$, of $m_{H_u}$ and $m_{H_d}$ in the two cases. In particular, in view of the results of the previous section, we study the mass of the $CP$-odd Higgs boson $A$. Notice that, though the differences in the soft scalar Higgs masses are significant, the value of $m_A$ is mainly determined by $M_{1/2}$: the $\approx50\%$ difference between the $SU(5)$ and $SO(10)$ soft scalar Higgs masses squeezes to a few percent correction in $m_A$. Henceforth, we expect the same funnel appearing in $SU(5)$ to take place also in the general $SO(10)$ case.\\ 
In fig.~\ref{so10cosmo} we show, with the same notation of fig.~\ref{su5cosmo}, the allowed parameter space in the $SO(10)$ case. The general features of the $SO(10)$ case are, as expected, remarkably similar to the $SU(5)$ case, thus confirming that the model-dependence of no-scale models with $M_{\rm bc}$ above $M_{\sss\rm GUT}$ is rather mild, as pointed out in \cite{minimalgaugino}. This weak dependence on the GUT structure which dictates the running above $M_{\sss\rm GUT}$ is easily understood: the values of the soft masses at $M_{\sss\rm GUT}$ are determined at one loop only by gauge charges through the non-vanishing gaugino masses. All other interactions are one-loop suppressed, and therefore only slightly affect the soft scalar mass pattern at the GUT scale. Moreover, the small splittings in the scalar SSB masses are partly washed out by the common MSSM running between the unification and the weak scales, dominated by gaugino masses. Nevertheless, it is somewhat non trivial that in $SO(10)$ as well the low energy spectrum allows the fulfillment of the $A$ Higgs resonance condition $2m_\chi\simeq m_A$, in a very similar range of $\tan\beta$ as in the $SU(5)$ case.\\
As regards the differences between the two considered GUT models, we point out that in the $SO(10)$ case the spectrum at the low energy scale is generically slightly {\em heavier} than in $SU(5)$, as emphasized before. This amounts to shifting the range of $\tan\beta$ and the lower bound on $M_{1/2}$ towards larger values. In the relic density as well we notice that the isolevel curves are closer to each other: this depends on the fact that in the $SO(10)$ case the variations of $m_A$ with $\tan\beta$ are larger than in $SU(5)$, hence the condition $2m_\chi\approx m_A$ is fulfilled in a smaller range of $\tan\beta$, and the funnel is slightly narrowed.\\ 
As far as the detectability of the model is concerned (lower panel), we draw the same conclusions as in the previous section: direct detection will be possible only at next to next generation experiments, and the large $m_{\chi}$ points on top of the $A$-pole funnel at $\tan\beta\simeq42$ will be beyond reach of CERN LHC\footnote{We are assuming here that the CERN LHC reach for general $SO(10)$ models is comparable to the one for $SU(5)$ models, as suggested by the highlighted strong similarities in the low energy spectra of the two cases.}.\\
As emerging from the pattern shown in the right panel of fig.~\ref{running} and from the preceding remarks, lowering $M_{\rm bc}$ would further reduce the differences between $SU(5)$ and $SO(10)$. Therefore we expect, in the framework of $SO(10)$, a scenario very similar to the one depicted in fig.~\ref{scale}, in the case $M_{\rm bc}<M_{\rm Pl}$.

\section{Conclusions}

To summarize, we showed that GUT models with no-scale boundary conditions above the unification scale and with $b$-$\tau$ Yukawa unification are compatible with the known cosmo-phenomenological constraints. In particular, the low energy spectrum of the two GUTs under scrutiny, $SU(5)$ and general $SO(10)$, allows, for $\tan\beta$ between 40 and 45, the fulfillment of the $A$-pole condition $2m_\chi\approx m_A$. The resulting enhancement of direct $s$-channel neutralino annihilations reduces the neutralino relic abundance $\Omega_\chi h^2$ within the current cosmologically preferred range. Further, we showed that when $M_{\rm bc}$ is around $10^{17}\ {\rm GeV}$ coannihilations between the neutralino and the lightest stau can also conspire to bring $\Omega_\chi h^2$ to sufficiently low values. We emphasize that, had it not been for the mentioned neutralino relic density suppression mechanisms, the models under scrutiny would have had no parameter space regions compatible with the current upper bound on $\Omega_{\sss\rm CDM} h^2$.\\ The analysis of the prospects for direct neutralino detection and for accelerator searches indicates that the first will be possible only at future experiments and in limited areas of the parameter space, while the latter will cover a large part, though not all, of the cosmologically and phenomenologically viable regions. Finally, a comparison between the two examined GUTs lead us to conclude that the dependence of no-scale models on the peculiar GUT structure is significantly mild.

\begin{acknowledgments}
The Author wishes to acknowledge P.~Ullio and C.~E.~Yaguna for valuable discussions and suggestions, and the Referee for useful comments and remarks.
\end{acknowledgments}

\end{document}